\newcommand{\un}{~\mathrm}

\documentclass[twocolumn
,floatfix
,showpacs
,preprintnumbers
,amsmath
,amssymb
,superscriptaddress]{revtex4}

\usepackage{graphicx}
\usepackage{dcolumn}
\usepackage{bm}

\begin{document}
\title{Scaling exponents for fracture surfaces in homogenous glass and glassy ceramics}
\author{D. Bonamy}
\author{L. Ponson}
\author{S. Prades}
\author{E. Bouchaud}
\author{C. Guillot}
\affiliation{Fracture Group, Service de Physique et Chimie des Surfaces et Interfaces, DSM/DRECAM/SPCSI, CEA Saclay,
F-91191 Gif sur Yvette, France} 

\begin{abstract}
We investigate the scaling properties of post-mortem fracture surfaces in silica glass and glassy ceramics. In both cases, the 2D height-height correlation function is found to obey Family-Viseck scaling properties, but with two sets of critical exponents, in particular a roughness exponent $\zeta\simeq 0.75$ in homogeneous glass and $\zeta\simeq 0.4$ in glassy ceramics. The ranges of length-scales over which these two scalings are observed are shown to be below and above the size of process zone respectively. A model derived from Linear Elastic Fracture Mechanics (LEFM) in the quasistatic approximation succeeds to reproduce the scaling exponents observed in glassy ceramics. The critical exponents observed in homogeneous glass are conjectured to reflect damage screening occurring for length-scales below the size of the process zone.
\end{abstract}
 
\date{\today}
\maketitle

The morphology of fracture surfaces is a signature of the complex damage and fracture processes occurring at the microstructure scale that lead to the failure of a given heterogeneous material. Since the pioneering work of Mandelbrot \cite{Mandelbrot84_nature}, a large amount of studies have shown that crack surface roughening exhibits some universal scaling features: Fracture surfaces were found to be self-affine over a wide range of length scales, characterized by a {\em universal} roughness exponent $\zeta \approx 0.8$, weakly dependent on the nature of the material and on the failure mode (see {\em e.g.} \cite{Bouchaud97_jpcm} for a review). Very recent studies \cite{Ponson06_prl} showed that a complete description of the scaling properties of fracture surfaces calls for the use of the {\em two-dimensional} (2D) height-height correlation function. This function was observed to exhibit anisotropic scaling properties similar to the Family-Viseck scaling \cite{Family91_book} predicted in interface growth models \cite{Barabasi95_book}, characterized by {\em three} critical exponents independent to some extent of the considered material, the loading condition and the crack growth velocity.

The origin of the scaling properties of fracture surfaces is still debated. Hansen and Schmittbuhl \cite{Hansen03_prl} suggested that the universal scaling properties of fracture surfaces are due to the fracture propagation being a damage coalescence process described by a stress-weighted percolation phenomenon in a self-generated quadratic damage gradient.  Bouchaud {\em et al.} \cite{Bouchaud93_prl} proposed to model the fracture surface as the trace left by a crack front moving through randomly distributed microstructural obstacles - the dynamics of which is described through a phenomenological nonlinear Langevin equation, keeping only the terms allowed by the symmetry of the system. Finally, Ramanathan {\em et al} used Linear Elastic Fracture Mechanics (LEFM) to derive a linear nonlocal Langevin equation within  both elastostatic \cite{Ramanathan97_prl} and elastodynamic \cite{Ramanathan97b_prl} approximation. All these approaches succeed to reproduce scale invariant crack surface roughness in qualitative - but unfortunately not quantitative - agreement with the experimental observations \cite{Ponson06_prl}.   

The universality of the roughness exponent was found to suffer from several exceptions: Metallic surfaces investigated at the nanometer scale were found to exhibit self-affine scaling properties, but with a roughness exponent significantly smaller than 0.8, closer to 0.4-0.5 \cite{Milman94_pms,Daguier97_prl}. This was first interpreted as a kinetic effect similar to the one expected for a moving line close to its depinning transition \cite{Daguier97_prl} - the small (resp.  large) scale roughness exponent 0.5 (resp. 0.8) corresponding to effective quenched noise (resp. thermal noise) \cite{Leschhorn97_ap}. The relevance of such an interpretation was later questioned since no small scale $\zeta\simeq 0.4-0.5$ roughness exponent was observed for nano-resolved fracture surface of Silica glass broken under stress corrosion with crack growth velocity as small as the picometer per second \cite{Ponson06_prl}. Furthermore, recent experiments reported similar values $\zeta\simeq 0.4-0.5$ {\em at large length-scales} in sandstone \cite{Boffa98_epjap}, artificial rock \cite{Bouchbinder05_prl} and glassy ceramics \cite{Ponson06b_prl}. In this latter case, the roughness exponent was found to be independent of the bead size, the porosity, the transgranular/intergranular nature of the failure mode and the crack growth velocity. This suggests the existence of a {\em second} universality class for failure problems.

The series of experiments reported here were designed to uncover the origin of these two distinct universality classes and focus more specifically on the range of {\em length-scales} over which the scaling properties are observed. Two materials are investigated: homogenous glass and glassy ceramics made of sintered $100~\mu\mathrm{m}$ glass beads. In both cases, the fracture surfaces are found to exhibit Family-Viseck scaling properties but with two different sets of critical exponents, in particular $\zeta\simeq 0.75$ for homogenous glass and $\zeta\simeq 0.4$ for glassy ceramics. The range of length-scales over which these two scalings are observed are shown to be below and above the size of process zone, respectively. Using LEFM, we show that the crack roughness development can been described as an elastic string with nonlocal interactions creeping in a 2D random medium - the spatial coordinate along which the crack globally grows playing the role of time. This approach allows to account {\em quantitatively} for the value of the observed critical exponents in the case of glassy ceramics. The role of damage in the case of homogeneous glass is finally discussed.  

\begin{figure}[!ht]
\includegraphics[width=0.7\columnwidth]{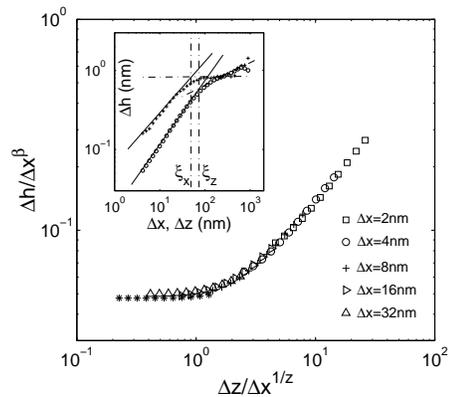}
\centering
\caption{Inset: 1D height-height correlation function calculated along the propagation direction $z$ and the crack front direction $x$ on a fracture surface of silica glass. In this experiment, the crack growth velocity was $v=10^{-10}\un{m/s}$. The straight lines are power law fits (see text for details). The vertical dot-dash line sets the cutoff lengths $\xi_z \simeq 70\un{nm}$ and $\xi_x\simeq 50\un{nm}$ of the self-affine regime in the $z$ and $x$ directions respectively. (b): The insets show the 2D height-height correlation functions $\Delta h_{\Delta x}(\Delta z)$ corresponding to different values of $\Delta x$ vs $\Delta z$. The data collapse was obtained using Eq. \ref{cor2D} with exponents $\zeta \simeq 0.75$, $\beta \simeq 0.6$, and $z =\zeta/\beta \simeq 1.2$.}
\label{fig1}
\end{figure}

{\em Experiments. - }
In all the following, the reference frame $(\vec{e}_x,\vec{e}_y,\vec{e}_z)$ is chosen so that $\vec{e}_x$, $\vec{e}_y$ and $\vec{e}_z$ are parallel to the propagation, loading and crack front directions respectively. Fracture surfaces in amorphous silica were obtained for various growth velocity ranging from $v=10^{-11}$ to $v=10^{-4}\un{m/s}$ using the procedure described in Ref. \cite{Prades05_ijss, Ponson06_prl}. Their topography was then measured through Atomic Force Microscopy with in-plane and out-of-plane resolutions estimated to be $5\un{nm}$ and $0.1\un{nm}$ respectively. The resulting images are $1024 \times 1024\un{pixels}^2$ and represent a square field of $1 \times 1~\mu\mathrm{m}^2$. The scaling properties of the fracture surfaces were analysed using the procedure discussed in Ref. \cite{Ponson06_prl}: First, the 1D height-height correlation function $\Delta h(\Delta z)=<(h(z+\Delta z,x)-h(z,x))^{2}>^{1/2}$ along the $z$ direction and  $\Delta h(\Delta x)=<(h(z,x+\Delta x)-h(z,x))^{2}>^{1/2}$ along the $x$ direction were computed. Both $\Delta h(\Delta z)$ and $\Delta h(\Delta x)$ were found to exhibit power-law behaviours, characterized by exponents $\zeta\simeq 0.75$ and $\beta\simeq 0.6$ respectively, extending up to length scales $\xi_z$ and $\xi_x$ respectively (Fig. \ref{fig1}: Inset). Second, the 2D height-height correlation function was computed (Fig. \ref{fig1}) and was shown to follow a Family-Viseck scaling \cite{Family91_book}:

\begin{equation}
\begin{array} {l}
\Delta h\propto\Delta x^{\beta}f(\frac{\Delta z}{\Delta x^{1/z}}),~	f(u) \sim \left\{
\begin{array}{l l}
1 & $if$ \quad u \ll 1  \\
u^{\zeta} & $if$ \quad u \gg 1
\end{array}
\right.
\end{array}
\label{cor2D}
\end{equation}

\noindent with $\zeta\simeq0.75$, $\beta\simeq0.6$ and $z=\zeta/\beta\simeq1.2$, for length-scales $\Delta z \leq \xi_z$ and $\Delta x \leq \xi_x$, in agreement with previous studies reported in \cite{Ponson06_prl}. These exponents $\zeta$, $\beta$ and $z$ were shown to correspond to the roughness, growth and dynamic exponents defined in interface growth problems. 

\begin{figure}[!ht]
\includegraphics[width=0.7\columnwidth]{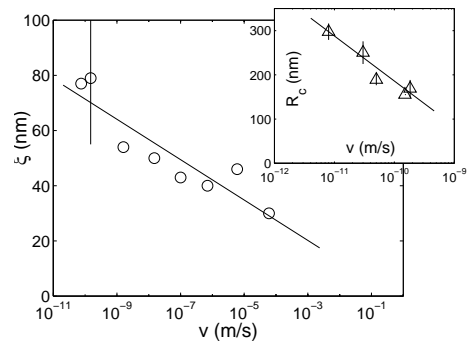}
\centering
\caption{Variation of the cutoff length $\xi$ (measured along $z$) as a function of the crack growth velocity $V$. The axes are semilogarithmic. The straight line corresponds to a fit $\xi\propto \log(v)$. Inset: Variation of the size of the damage zone $R_c$ (measured along $x$) as a function of the crack growth velocity $v$ (see text for details). The axes are semilogarithmic. The straight line correspond to a fit $R_c\propto \log(v)$ }
\label{fig2}
\end{figure}

Then, we investigated the influence of the crack growth velocity $v$ on the scaling properties of the post-mortem fracture surface. The critical exponents $\zeta$, $\beta$ and $z$ do not show any noticeable dependence on $v$. On the other hand, the cutoff length $\xi$ was observed to decrease slowly, as the logarithm of $v$ (Fig. \ref{fig2}). For the smallest value of $v$, ranging from $10^{-11}$ to $10^{-9}\un{m/s}$, we were able to observe in real-time, at the nanometer scale, the crack propagation during the specimen failure \cite{Prades05_ijss}. At these scales, it was shown that the deformation fields does not fit with the linear elastic predictions over a fairly large region ($\sim 100$ nanometers) at the crack tip \cite{Guilloteau97_epl,Celarie03_prl,Prades05_ijss}. This zone is thereafter referred to as the damage zone or the process zone. The variation of its size $R_c$ with respect to the crack velocity $v$ are presented in the inset of Fig. \ref{fig2}. First, $R_c$ is found to be larger, but of the same order of magnitude, than $\xi$. Second, $R_c$, like $\xi$, is observed to decrease as the logarithm of $v$. This leads us to conjecture that the damage zone size $R_c$ is the relevant length-scale that set the upper cutoff length $\xi$ that limits the scaling given by Eq. \ref{cor2D} with the exponents $\{\zeta\simeq0.75,\beta\simeq0.6,z=\zeta/\beta\simeq1.2\}$. At these length-scales, the material cannot be identified with a coarse-grained equivalent linear elastic medium, which explains the failure of existing models \cite{Ramanathan97_prl} derived from Linear Elastic Fracture Mechanics (LEFM) to reproduce the critical exponents $\{\zeta\simeq0.75,\beta\simeq0.6,z=\zeta/\beta\simeq1.2\}$ observed experimentally.  

\begin{figure}[!ht]
\includegraphics[width=0.7\columnwidth]{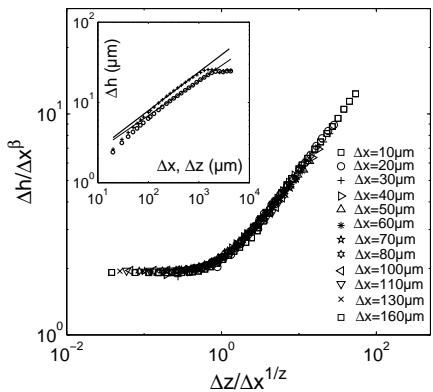}
\centering
\caption{The insets show the 2D height-height correlation functions $\Delta h_{\Delta x}(\Delta z)$ corresponding to different values of $\Delta x$ vs $\Delta z$ for a fracture surface of glassy ceramic made of glass beads with a porosity $\Phi = 6\%$. The data collapse was obtained using Eq. \ref{cor2D} with exponents $\zeta \simeq 0.4$, $\beta \simeq 0.5$, and $z =\zeta/\beta\simeq 0.8$.}
\label{fig3}
\end{figure}

We now examine the scaling properties of fracture surfaces in glassy ceramics made of sintered beads of Silicate glass with diameter $d$ ranging between $104-128~\mu\mathrm{m}$ (see Ref. \cite{Ponson06_prl} for details). In this class of materials, the size of the process zone observed in the vicinity of the various (micro-)crack tips are expected to be of the order of $100\un{nm}$ as in homogeneous glass, while the microstructure scale is set by the mean bead diameter at a length-scale three order of magnitudes larger. Figure \ref{fig3} presents the 2D height-height correlation function. As for homogeneous glass, this function is found to follow the Family-Viseck scaling \cite{Family91_book} given by Eq. \ref{cor2D}, but with a different set of critical exponents $\{\zeta\simeq0.4,\beta\simeq0.5,z=\zeta/\beta\simeq0.8\}$. This set of exponents was found to be independent of the porosity, the bead diameter and the crack growth velocity.

{\em Intepretation.-}
Since the scaling properties of glassy ceramics were observed at length-scales {\em larger than} the process zone size,it is natural to attempt to interpret these scaling properties within LEFM framework. We restrict the following analysis to the case where (i) the elastic properties can be considered as homogeneous and (ii) the crack speed is small compared to the speed of wave propagation so that quasi-static approximation is relevant. In an isotropic elastic material, near the crack front, the stress field can be written \cite{Irwin58_proc}:

$\sigma_{ij}=\sqrt{\frac{1}{2 \pi r}}\{ K_I\Sigma^{I}_{ij}(\theta) +K_{II}\Sigma^{II}_{ij}(\theta) +K_{III}\Sigma^{III}_{ij}(\theta) \}$,

\noindent where $r$ is the distance from the crack front, $\theta$ is the angle with respect to the local direction of crack propagation, $\Sigma_{ij}$ are universal functions of $\theta$, and $K_I$, $K_{II}$ and $K_{III}$ are the mode I (tension), mode II (shear) and mode III (tear) intensity factors respectively. In an ideal homogeneous elastic material under tensile $K^0_I$ loading, the crack front would remain straight and would propagate within a given plane. 
But heterogeneities of a disordered material like glassy ceramics induce both in-plane distortion $f(z,t)$ and out-of-plane distortion $h(z,x=f(z,t))$ of the crack front. In turn, these deviations from straightness induce perturbations $\delta K_I$, $\delta K_{II}$ and $\delta K_{III}$ in the local loading of the crack front. To linear order, $\delta K_I$ depends only on $f$ \cite{Gao89_jam} while $\delta K_{II}$ and $\delta K_{III}$ are functional of $h$ only \cite{Larralde95_epl}. Since we are primarily interested in the fracture surfaces, we  focus on the interrelations between $\delta K_{II}$ and $\delta K_{III}$ and $h$, returning in the end of the paper to a brief discussion of the in-plane crack roughness.

The path chosen by a crack propagating in an elastic isotropic material is the one for which the local stress field at the tip is of mode I type ("criterion of local symmetry") \cite{Goldstein74_ijf,Cotterell80_ijf,Hodgdon93_prb}. In other words, the net mode II stress intensity factor should vanish in each location $z$ along the crack front. Three contributions should be taken into account in the evaluation of $\delta K_ {II}$. The first contribution is due to both inevitable imperfections in the loading system or in crack alignment and the heterogeneous nature of the material. This contribution is modelled by a quenched uncorrelated random field $\delta K_{II} ^{(1)} = K_I^0\eta(z,x,h)$ written, for sake of simplicity, as the sum of two uncorrelated random fields $\delta K_{II} ^{(1)} = K_I^0\{\eta_q(z,h)+\eta_t(z,x)\}$. The second contribution $\delta K_{II}^{(2)}$ arises from the coupling of the singular mode I component of the stress field of the unperturbed crack with the position of the crack edge. In Fourier space this contribution is given to linear order by \cite{Larralde95_epl}: $\delta \hat{K}_{II}^{(2)}(k_z,x)=K_I^0 \partial_x\hat{h} +\frac{1}{2}|k_z|K_I^0\frac{2-3\nu}{2-\nu}\hat{h}(k_z,x)$ where $\nu$ refers to the Poisson's ratio, and $\delta\hat{K}_{II}^{(2)}(k_z,x)$ (resp. $\hat{h}(k_z,x)$) refers to the Fourier transform of $\delta K_{II}^{(2)}(z,x)$ (resp. $h(z,x)$) with respect to $z$. Finally, a third contribution comes from the coupling between the slope of the crack surface and the non singular $T$ normal stress in the direction of crack propagation \cite{Cotterell80_ijf,Movchan98_ijss}. This third contribution was shown to be negligible with respect to the second contribution in the thermodynamic limit \cite{Larralde95_epl}. Finally, making the net mode II stress intensity factor vanish at each location $z$ leads to:

\begin{equation}
\frac{\partial h}{\partial x}=-J(z,x,\{h\})+\eta_q(z,h)+\eta_t(z,x)
\label{eqfinal}
\end{equation}

\noindent where the Fourier transformed elastic kernel $\hat{J}(k_z,x,\{\hat{h}\})$ is given by: 

\begin{equation}
\hat{J}(k_z,x,\{\hat{h}\})=|k_z|\frac{2-3\nu}{2-\nu}\hat{h}(k_z,x)
\label{kernel}
\end{equation}

\noindent In other words, the morphology of the fracture surface $h(x,z)$ is given by the motion of the elastic string $h(z)$ that "creeps" - the $x$ coordinate playing the role of time - within a random potential $\eta_q(z,h)$ due to the "thermal" fluctuations $\eta_t(z,x)$. The scaling properties of the surface $h(z,x)$ in the steady regime are then expected to be described by a 2D height-height correlation given by Eq. \ref{cor2D} \cite{Chauve00_prb}. Furthermore, if we consider an interface whose interaction kernel in momentum space scales as $J(k_z,\{\hat{h}\})=J_0 |k_z|^\mu\hat{h}$, the values of the critical exponents $\zeta$, $\beta$ and $z$ depends only on $\mu$. In particular, for long-range interaction $\mu=1$,  one gets $\zeta \simeq 0.39$ \cite{Schmittbuhl95_prl,Tanguy98_pre,Rosso02_pre}, $z \simeq 0.75$ \cite{Tanguy98_pre,Schmittbuhl95_prl} and $\beta=\zeta/z\simeq 0.5$ in perfect agreement with the values measured experimentally in glassy ceramics.

The values of the critical exponents experimentally observed in homogeneous silica glass can now be discussed. In this later case, the scaling properties were observed at length-scales {\em smaller} than the size of the process zone, {\em i.e.} at length-scales where the material cannot be considered as a linear elastic anymore. Recent AFM experiments \cite{Celarie03_prl,Prades05_ijss} and Molecular Dynamics (MD) observations \cite{vanBrutzel02_mrssp} have shown that damage spreading within this process zone occurs through the nucleation of nanoscale cavities whose nature remains controversial: They were first conjectured to be similar to the ones classically observed - at much larger scale - during the ductile fracture of metallic alloys \cite{Celarie03_prl,Prades05_ijss}. This interpretation was later questioned since these cavities were shown to leave no visible remnants on the post-mortem fracture surfaces \cite{Guin04_prl}. It appears then natural to conjecture that damage - independently of its precise nature - {\em screens} the elastic interactions $J(z,\{h\})$ within the process zone, making the effective $\mu$ {\em larger} than the value $\mu=1$ expected for perfectly linear elastic materials. Renormalisation Groups (RG) methods \cite{Leschhorn97_ap,Narayan93_prb} predict $\zeta = (2\mu-1)/3$, $z = (5\mu+2)/9$ to first order in $\epsilon=2\mu-1$. "Arbitrary" values $\mu \simeq 1.5-1.7$ would then allow us to account for the values of $\zeta\simeq0.75$ and $z \simeq 1.2$ observed in homogenous Silica glass, as well as for a wide range of materials \cite{Bouchaud97_jpcm,Ponson06_prl}. This has been confirmed through numerical simulations. Understanding how damage screening can select such an effective interaction range in crack problems provides a significant challenge for future investigation.

More generally, we conjecture that both critical scaling regimes can be observed in all the heterogeneous materials: For length-scales smaller (resp. larger) than the damage zone size, ones expects Family-Viseck scaling with $\{\zeta\simeq 0.75,\beta\simeq 0.6,z=\zeta/\beta\simeq1.2\}$ (resp. $\{\zeta\simeq 0.4,\beta\simeq 0.5,z=\zeta/\beta\simeq0.8\}$). 

It should be mentioned that LEFM applied to describe the motion of the in-plane crack front in a disordered material results in a Langevin equation with non-local elastic kernel and quenched noise \cite{Ramanathan97_prl,Schmittbuhl95_prl}. At the depinning transition, this approach predicts self-affine in-plane roughness characterised by a roughness exponent $\zeta\simeq 0.39$ \cite{Rosso02_pre} and a dynamic exponent $z\simeq 0.75$ \cite{Tanguy98_pre,Schmittbuhl95_prl}, while experiments \cite{Maloy97_prl} report values $\zeta\simeq 0.6$ and $z\simeq 1$. These experimental values are much closer to the ones expected in elastic line models with short range elastic interactions, that predict roughness exponents $\zeta\simeq 0.63$ and $z=1$ \cite{Rosso01_prl}, which suggests similar damage screening effects as the ones invoked for out-of-plane crack roughness.

Finally, it is worth noting that, to our knowledge, this study reports the first experimental observation of the critical roughnening predicted by the linear Langevin equation with non-local elastic interactions initially proposed to describe a broad variety of systems ranging from interfaces in disordered magnets \cite{Durin00_prl}, contact lines of liquid menisci on a rough substrate \cite{Joanny84_jcp}, and interfacial cracks in solids \cite{Schmittbuhl95_prl}. Fracture surfaces may thus represent an ideal experimental tool to investigate  such pinning/depinning physics.  Work in this direction is currently under progress.

\begin{acknowledgments}
We  gratefully acknowledge S. Roux for a critical reading of the manuscript. We thank T. Bernard and G. Le Chevallier for technical support, and P. Vi{\'e} for the making of the glassy ceramics. We are grateful to H. Auradou, J.-P. Bouchaud, K. Dahmen, O. Duemmer, J.-P. Hulin, I. Procaccia,  M. Robbins, C. L. Rountree, A. Rosso and S. Roux for many enlightening discussions.
\end{acknowledgments}

\end{document}